\documentclass[11pt, reqno]{amsart}
\usepackage{geometry}
\usepackage{graphicx}
\usepackage{amssymb}
\usepackage{amsmath, mathrsfs, stmaryrd}
\usepackage{comment, tensor}
\usepackage{color}
\newtheorem{theorem}{Theorem}
\newtheorem{definition}[theorem]{Definition}

\newtheorem{proposition}[theorem]{Proposition}

\newtheorem{question}[theorem]{Question}
\newcommand{\R}{\mathbb R}

\usepackage{xpatch}
\usepackage{hyperref}
\usepackage{footnotebackref}

\makeatletter
\xpretocmd{\@adminfootnotes}{\let\@makefntext\BHFN@OldMakefntext}{}{}
\renewcommand\@makefntext[1]{%
  \@ifundefined{@makefnmark}
    {}
    {%
     \renewcommand\@makefnmark{%
       \mbox{%
         \textsuperscript{%
           \normalfont
           \hyperref[\BackrefFootnoteTag]{\@thefnmark}%
         }%
       }\,%
     }%
     \BHFN@OldMakefntext{#1}%
  }%
}
\makeatother

\numberwithin{theorem}{section}
\numberwithin{equation}{section}

\title[angular momentum at infinity]{Limits of quasi-local angular momentum on an isolated gravitating system}

\author{Mu-Tao Wang}

\begin{document}
\begin{abstract}
I shall discuss the Chen-Wang-Yau quasilocal angular momentum, which is defined based on the theory of optimal isometric embedding and quasilocal mass of Wang-Yau, and the limits of which at  spatial and null infinity of an isolated gravitating system. This is based on joint work  with Po-Ning Chen, Jordan Keller, Ye-Kai Wang, and Shing-Tung Yau. 
\end{abstract}

\thanks{This material is based upon work supported by the National Science
Foundation under Grant Number DMS 1810856 (Mu-Tao\ Wang).  The author would like to thank Po-Ning, Chen, Jordan Keller, and Ye-Kai Wang for helpful discussions.}\maketitle
\section{Introduction}

In the theory of general relativity, the definition of angular momentum is proved to be a more challenging task than the definition of energy/mass. An essential difficulty for the definition of energy/mass goes back to Einstein's equivalence principle, which implies that gravitation, unlike other physical fields, has no density. Moreover, in most attempts in which the Hamiltonian method was employed, the issue is complicated by the fact that general relativity is a nonlinear theory and there is no canonical choice of a reference system. An isolated gravitating system corresponds to an asymptotically flat spacetime where gravitation is weak at infinity. In terms of  the asymptotically flat coordinate system, there are well-defined notions of energy/mass, most notably the ADM energy/mass at spatial infinity and the Bondi energy/mass at null infinity. However, the definitions of angular momentum at both spatial infinity and null infinity are more subtle due to the nature of asymptotically rotation Killing field and no consensus has been reached. 

A possible approach to define angular momentum at both spatial infinity and null infinity in a uniform way is to take the limit of a quasilocal definition. In this article, we first discuss the theory of quasilocal mass and optimal isometric embedding proposed by Wang-Yau in \cite{Wang-Yau1,Wang-Yau2} and the definition of Chen-Wang-Yau quasilocal angular momentum in \cite{Chen-Wang-Yau3, Chen-Wang-Yau4}. We then explain how the limits of the Chen-Wang-Yau quasilocal angular momentum at spatial and null infinity provide viable definitions of angular momentum for an isolated gravitating system.

\section{Wang-Yau quasilocal mass and Chen-Wang-Yau quasilocal angular momentum}
 The notion of quasilocal mass is attached to a 2-dimensional closed surface $\Sigma$ which bounds a spacelike region in spacetime. $\Sigma$ is assumed to be a topological 2-sphere, but with different intrinsic geometry and extrinsic geometry, we expect to read off the effect of gravitation in the spacetime vicinity of the surface. Suppose the surface is spacelike, i.e. the induced metric $\sigma$ is Riemannian.  An essential part of the extrinsic geometry is measured by the mean curvature vector field $\bf{H}$ of $\Sigma$. $\bf{H}$ is a normal vector field of the surface such that the null expansion along any null normal direction $\ell$ is given by the pairing $\langle {\bf H}, \ell\rangle$ of
$\bf{H}$ and $\ell$.

In \cite{Wang-Yau1}, Wang-Yau proposed the following definition of quasilocal mass which depends only on $\sigma$ and $\bf{H}$ of a 2-surface $\Sigma$ in spacetime.  To evaluate the quasilocal mass of $\Sigma$ with the physical data $(\sigma, \bf{H})$, one first solves the optimal isometric embedding equation, see \eqref{oiee} below, which gives an embedding of $\Sigma$ into the Minkowski spacetime with the image
surface $\Sigma_0$ that has the same 
induced metric as $\Sigma$, i.e. $\sigma$. One then compares the extrinsic geometries of $\Sigma$ and $\Sigma_0$ and evaluates the quasilocal mass from $\sigma, \bf{H}$ and $\bf{H_0}$.

Assuming the mean curvature vector ${\bf H}$ is spacelike, the physical surface $\Sigma$ with physical data $(\sigma, \bf{H})$ gives $(\sigma, |\bf{H}|, \alpha_{\bf {H}})$ where $|{\bf H}|>0$ is the Lorentz norm of $\bf{H}$ and $\alpha_{\bf H}$ is the connection one-form determined by $\bf{H}$. Given an isometric embedding $X:\Sigma\rightarrow \R^{3,1}$ of $\sigma$. Let $\Sigma_0$ be the image $X(\Sigma)$ and $(\sigma, |\bf{H}_0|, \alpha_{\bf {H}_0})$ be the corresponding data of $\Sigma_0$ (${\bf H_0}$ is again assumed to be spacelike).

Let $T$ be a future timelike unit Killing field of $\R^{3,1}$ and define $\tau=-\langle X, T\rangle$ as a function on $\Sigma$. Define a function $\rho$ and a 1-form $j_a$ on $\Sigma$:
  \[ \begin{split}\rho &= \frac{\sqrt{|{\bf H}_0|^2 +\frac{(\Delta \tau)^2}{1+ |\nabla \tau|^2}} - \sqrt{|{\bf H}|^2 +\frac{(\Delta \tau)^2}{1+ |\nabla \tau|^2}} }{ \sqrt{1+ |\nabla \tau|^2}}\\
 j_a&=\rho {\nabla_a \tau }- \nabla_a \left( \sinh^{-1} (\frac{\rho\Delta \tau }{|{\bf H}_0||{\bf H}|})\right)-(\alpha_{{\bf H}_0})_a + (\alpha_{{\bf H}})_a, \end{split}\] where $\nabla_a$ is the covariant derivative with respect to the metric $\sigma$, $|\nabla \tau|^2=\nabla^a \tau\nabla_a \tau$ and $\Delta \tau=\nabla^a\nabla_a \tau$. 
$\rho$ is the quasilocal mass density and $j_a$ is the quasilocal momentum density. A full set of quasilocal conserved quantities was defined in \cite{Chen-Wang-Yau3, Chen-Wang-Yau4} using $\rho$ and $j_a$. 
 
The optimal isometric embedding equation for $(X, T)$ is 
\begin{equation}\label{oiee} \begin{cases}
\langle dX, dX\rangle&=\sigma\\
\nabla^a j_a&=0.
\end{cases}\end{equation}
The first equation is the isometric embedding equation into the Minkowski spacetime and the second one is the Euler-Lagrange equation of the quasilocal energy $E(\Sigma, \tau)$ \cite{Wang-Yau1,Wang-Yau2} in the space of isometric embeddings.
The quasi-local mass for the optimal isometric embedding $(X, T)$ is defined to be \[E(\Sigma, X, T)=\frac{1}{8\pi}\int_\Sigma \rho.\]
 It is shown in \cite{Wang-Yau1, Wang-Yau2} that $E(\Sigma, X, T)$ is { positive in general}, and { zero for surfaces in the Minkowski spacetime}. 

The theory of quasilocal mass and optimal isometric embedding was employed by Chen-Wang-Yau in \cite{Chen-Wang-Yau3, Chen-Wang-Yau4} to define quasilocal conserved quantities. For an optimal isometric embedding $(X, T)$, by restricting a rotation (or boost) Killing field $K$ of $\R^{3,1}$ to $\Sigma_0=X(\Sigma)\subset \R^{3,1}$, the quasi-local conserved quantity is defined to be: 
\[-\frac{1}{8\pi} \int_\Sigma \langle K, T\rangle \rho+(K^\top)^a  j_a ,\] where $K^\top$ is the component of $K$ that is tangential to $\Sigma_0$.
In particular, $K=x^i\partial_j-x^j \partial_i, i<j$ defines an angular momentum with respect to $\partial_t$. Here $(t, x^i)$ and $(\partial_t, \partial_i)$ are standard coordinates and coordinate vectors of the Minkowski spacetime.

The image of the optimal isometric embedding $\Sigma_0$ is essentially the ``unique" surface in the Minkowski spacetime that best matches the physical surface $\Sigma$. 
If the original surface $\Sigma$ happens to be a surface in the Minkowski spacetime, the above procedure identifies $\Sigma_0=\Sigma$ up to a global isometry. 

A solution of the optimal isometric embedding equation is indeed a critical point of the quasilocal energy $E(\Sigma, \tau)$. In \cite{Chen-Wang-Yau2}, we study the minimizing and uniqueness property for a solution of the optimal isometric embedding equation. In particular, the following theorems hold true:

\begin{theorem}\cite{Chen-Wang-Yau2}
Let $(\sigma, H)$ be the data of a spacelike surface $\Sigma$ with spacelike mean curvature vector $H$ in the Minkowski spacetime and $T$ be a unit timelike Killing field. Suppose the projection of $\Sigma$ onto the orthogonal complement of $T$ is a convex surface. Then

(1) the kernel of the linearized optimal isometric 
embedding system consists precisely of Lorentz transformations. 

(2) the second variation of the quasilocal energy  $E(\Sigma, \tau)$
is non-negative definite. \end{theorem}

For a spacelike surface with spacelike mean curvature vector in a general spacetime, one has

 \begin{theorem} \cite{Chen-Wang-Yau2} Let $(\sigma, H)$ be the data of a spacelike surface $\Sigma$ in a general spacetime. Suppose that $\tau_0$ is a critical point of the quasi-local energy $E(\Sigma, \tau)$ and that the corresponding quasilocal mass density $\rho$ is positive, then $\tau_0$ is a local minimum for  $E(\Sigma,\tau)$.
\end{theorem}

The case when the reference isometric embedding lies in a totally geodesic spacelike subspace of the Minkowski spacetime 
 and thus $\tau_0=0$ was proved by Miao-Tam-Xie \cite{mtx}.
The proof of the general case in \cite{Chen-Wang-Yau2} consists of a non-linear comparison principle $E(\Sigma,\tau) \ge E(\Sigma,\tau_0) +E(\Sigma_{\tau_0}, \tau)$ and the identification of the equality case of the positive energy theorem. 

The above theorems allow us to solve the optimal isometric embedding system for configurations that limit
to a surface in the Minkowski spacetime. This is in particular sufficient for calculations at infinity of an isolated system when the total 
mass is positive.

%%%%%%%%%%%%%%%%%%%%%%%%%%%%%%%%%%%%%%%%%%%%%%%%%%%%%%%%%%%%%%
\section{Angular momentum at spatial infinity}

\subsection{The ADM angular momentum}
Let $(M, g)$ be a Riemannian 3-manifold and $k$ be a symmetric 2-tenor on $M$. 
 Recall that $(M,g,k)$ is said to be an  asymptotically flat initial data set if (1) 
there exists a compact subset $K$ of $M$ such that $M\backslash K$ is diffeomorphic to a finite union of ends $\cup_i  \mathbb{R}^3\backslash B_i$ where  each $B_i$ is a geodesic ball in $\mathbb{R}^3$, and (2) there exists an asymptotically
flat coordinate system $(x^1, x^2, x^3)$ on each end, such that 
\begin{equation} \label{decay} g=\delta+O_2(r^{-q}) \text{ and } k=O_1(r^{-p}),\end{equation}  where  $r=\sqrt{\sum_{i=1}^3 (x^i)^2}$ and \begin{equation}\label{af}q>\frac{1}{2} \text{ and } p>\frac{3}{2}.\end{equation} The ADM (Arnowitt-Deser-Misner) mass \cite{Arnowitt-Deser-Misner} of $(M, g, k)$ is defined to be:
\[\frac{1}{16\pi} \int_{S^2_\infty} \sum_{i, j} (g_{ij, j}-g_{jj,i}) \nu^i,\] where $S^2_\infty$ is the limit as $r\rightarrow \infty$ of coordinate spheres $S_r$ and $\nu=\nu^i\partial_i$ is the outward unit normal of $S_r$.   

The decay rate assumption on the asymptotically flatness \eqref{af} is crucial. Under this assumption, the ADM mass satisfies the invariance property \cite{Bartnik1986} and the important positivity and rigidity property by the positive mass theorem of Schoen-Yau and Witten \cite{SY1, SY2, W}. 
  
In addition to the ADM mass and ADM energy-momentum, there is also a companion definition of angular momentum that is also attributed to ADM (Arnowitt-Deser-Misner) and is defined as  \[J=\frac{1}{8\pi} \int_{S^2_\infty} \pi(x^i\partial_j-x^j\partial_i, \nu), i<j, \text{ where } \pi=k-(tr_g k)g, \] where $x^i\partial_j-x^j\partial_i$ is considered to be an asymptotically rotation Killing field with respect to the asymptotically flat coordinate system $(x^1, x^2, x^3)$. 
However, the calculation of angular momentum is more subtle, as the expression diverges apparently. There are proposals by Regge-Teitelboim \cite{RT} of a parity condition on $(g, k)$ to ensure finiteness, and important gluing constructions
and density theorems  for prescribing angular momentum by Corvino-Schoen \cite{CS2006}, Chru\'sciel-Delay \cite{Chrusciel-Delay-03}, Chru\'sciel-Corvino-Isenberg \cite{Chrusciel-Corvino-Isenberg-11}, Huang-Schoen-Wang \cite{Huang-Schoen-Wang:2011} etc. under such a condition.

Without the parity condition, it was observed by Chru\'sciel \cite{Ch_1987_angular} that the ADM angular momentum is finite if $p+q>3$ in \eqref{decay}.
However, in \cite{CHWY}, it was shown that there exist asymptotically flat spacelike hypersurfaces in the Minkowski or Schwarzschild spacetime with finite, nonzero ADM angular momentum such that $g=\delta+O(r^{-\frac{4}{3}})$ and $k=O(r^{-\frac{5}{3}})$.  Theses hypersurfaces still have the expected ADM mass as the asymptotically flat conditions $q>\frac{1}{2}$ and $p>\frac{3}{2}$ are still satisfied. However, it is difficult to interpret the nonzero angular momentum when the corresponding spacetime is static. It is thus of interest to investigate to what extent is the ADM definition a valid one. In particular, one can ask the following question:
\begin{question}\label{Kerr} For an asymptotically flat spacelike hypersurface in the Kerr spacetime with $g=\delta+O(r^{-q})$ and $k=O(r^{-p})$ such that  $q>\frac{4}{3}$ and $p>\frac{5}{3}$, is the ADM angular momentum the expected one?\end{question}

\subsection{The limit of CWY quasilocal angular momentum at spatial infinity}

Before the Chen-Wang-Yau (CWY) definition, there were several proposals of the definition of quasilocal angular momentum. Most notably, in the axi-symmetric case, there was the Komar angular momentum. For a general spacetime, there were definitions proposed by Penrose \cite{Penrose1}, Dougan-Mason \cite{DM}, Ludvigsen-Vickers \cite{LV} etc, which are based on twistor or spinor constructions. 

 A key question is what justifies a good definition of angular momentum at the quasilocal level. For the definition of quasilocal mass, obviously one requires that the quasilocal mass should be positive in general and should be zero for a surface in the Minkowski spacetime \cite{Christodoulou-Yau-88}, as well as that the limit at spatial infinity should recover the ADM mass. 
 All previous known definitions of quasilocal angular momentum satisfy the covariant properties with respect to the Poincare group and consistency with the Komar definition. 
 Each quasilocal definition of angular momentum gives a limit at spatial infinity which can be viewed as the  total angular momentum of an asymptotically flat initial data set.
 A natural criterion is thus to investigate these limits and see if they provide a viable definition. Especially one sees from last section that  there are difficulties for the ADM definition in the Minkowski or Schwarzschild spacetime.

The limit of CWY quasilocal angular momentum gives a definition total angular momentum in the following way. Given an asymptotically flat coordinate system on an end of an asymptotically flat initial data set $(M, g, k)$, consider the coordinate sphere $S_r$. 
Suppose the ADM mass of $(M, g, k)$ is positive, then there is a unique, locally energy-minimizing, optimal isometric embedding of $S_r$ whose image approaches a large round sphere in $\R^3$. Take the limit as $r\rightarrow \infty$ of the quasi-local conserved quantities on $S_r$, we obtain $(E, P_i, J_i, C_i)$ where $(E, P_i)$ is the same as the ADM energy-momentum \cite{Chen-Wang-Yau1}.

An invariance theorem in the Kerr spacetime was proved in \cite{Chen-Wang-Yau3}: any ``strictly spacelike" hypersurface in the Kerr spacetime has the same total angular momentum (Cf. Question \ref{Kerr}). Here ``strictly spacelike" means, in Boyer-Lindquist coordinates $(t, r, \theta, \phi)$, outside a compact subset the hypersurface is given by $t=O(c r)$ for a constant $c$ with $|c|<1$. In particular, the CWY angular momentum differs from the ADM definition, as it vanishes for these hypersurfaces in the Minkowski or the Schwarzschild spacetime. The proof relies on a gravitational conservation law. The CWY definition is also conserved along the vacuum Einstein equation \cite{Chen-Wang-Yau3, Chen-Wang-Yau4}.

 In addition, the limit of quasilocal conserved quantities for spacelike hypersurface of harmonic asymptotics of Corvino-Schoen were computed in \cite{CW}.

\section{Angular momentum at null infinity}
In this section, we focus on the definition of total angular momentum at future null infinity $\mathfrak{I}^+$. There were various definitions and proposals of total angular momentum at future null infinity $\mathscr{I}^+$ that include, but not limit to,   Ashtekar-Hansen \cite{AH}, Barnich-Troessaert \cite{BT}, Bramson\cite{Bramson}, Chru\'sciel-Jezierski-Kijowski \cite{CJK}, Dougan-Mason \cite{DM}, Dray-Streubel \cite{DS}, Hawking-Perry-Strominger \cite{HPS}, Ludvigsen-Vickers \cite{LV}, Rizzi \cite{Rizzi}, Winicour-Tamburino \cite{WT}, etc.

These definition relies on choice of coordinate system or guage and choice of reference swhich determines the asymptotically rotation Killing field. Again, a key issue here is to identify good criteria to justify these definitions. We emphasize on the invariance/equivariance property, especially with respect to the BMS group. 
We will first review the well-known description of null infinity in terms of the Bondi-Sachs coordinate system and the invariance/equivariance property of the Bondi-Sachs energy-momentum. At the end, we discuss how the limit of the CWY quasilocal angular momentum gives an angular momentum definition at null infinity.

\subsection{The description of null infinity}
 
The spacetime near $\mathscr{I}^+$ is described in terms of the Bondi-Sachs coordinates $(u, r, \theta, \phi)$ which are chosen in the following way.  Level sets of $u$ are null hypersurfaces generated by null geodesics, $\theta$ and $\phi$ are extended  by constancy along the integral curves of the gradient vector field of $u$, and 
 $r$ corresponds to the ``area distance".
 %\begin{center}
%{\includegraphics[height=5cm]{P2.PNG}}
%\end{center}
In terms of a Bondi-Sachs coordinate system $(u, r,  x^2, x^3)$, the spacetime metric takes the form
\begin{equation}\label{spacetime_metric}g_{\alpha\beta}dx^\alpha dx^\beta= -UV du^2-2U dudr+r^2 h_{ab}(dx^a+W^a du)(dx^b+W^b du).\end{equation} The index conventions here are $\alpha, \beta=0,1, 2, 3$, $a, b=2, 3$, and $u=x^0, r=x^1$. See \cite{CMS, MW} for more details of the construction of the coordinate system. 
The metric coefficients $U, V, h_{ab}, W^a$  of \eqref{spacetime_metric} depend on $u, r, \theta, \phi$, but the assumption on $r$ implies that the determinant condition that $\det h_{ab}$ is independent of $u$ and $r$. These gauge conditions reduce the number of metric coefficients of a Bondi-Sachs coordinate system to six (there are only two independent components in $h_{ab}$). On the other hand, the boundary conditions $U\rightarrow 1$, $V\rightarrow 1$, $W^a\rightarrow 0$, $h_{ab}\rightarrow \sigma_{ab}$ are imposed as $r\rightarrow \infty$ (such boundary conditions may not be satisfied in a radiative spacetime).   The special gauge choice of the Bondi-Sachs coordinates implies a hierarchy among the vacuum Einstein equations, see \cite{MW, HPS}.

 Assuming the outgoing radiation condition \cite{Sachs, VDB, MW, Winicour, VK}, the boundary condition and the vacuum Einstein equation imply that as $r\rightarrow \infty$, all metric coefficients can be expanded in inverse integral powers of $r$.\footnote{The outgoing radiation condition assumes the traceless part of the $r^{-2}$ term in the expansion of $h_{ab}$ is zero. The presence of this traceless term will lead to a logarithmic term in the expansions of $W^a$ and $V$. Spacetimes with metrics which admit an expansion in terms of $r^{-j}\log^i r$ are called ``polyhomogeneous" and are studied in \cite{CMS}. They do not obey the outgoing radiation condition or the peeling theorem \cite{VK}, but they do appear as perturbations of the Minkowski spacetime by the work of Christodoulou-Klainerman \cite{CK}.} In particular, 
\[\begin{split} U&=1+O(r^{-2}),\\
V&=1-\frac{2m}{r}+O(r^{-2}),\\
W^a&=O(r^{-2}),\\
h_{ab}&={\sigma}_{ab}+\frac{C_{ab}}{r}+O(r^{-2})\end{split},\] where  $m=m(u, x^a)$ is the mass aspect and $C_{ab}=C_{ab}(u, x^a)$ is the shear tensor of this Bondi-Sachs coordinate system.
The Bondi-Sachs  energy-momentum 4-vector associated with a $u$-slice is then \[e(u)=\frac{1}{4\pi}\int_{S^2_\infty} m(u, x^a) dv_\sigma,\,\, p_i(u)=\frac{1}{4\pi}\int_{S^2_\infty} m(u, x^a) Y_i dv_\sigma, i=1,2, 3\] where $\{Y_i=Y_i(x^a), i=1, 2, 3\}$ is an orthonormal basis of the $(-2)$ eigenspace of $\Delta=\Delta_\sigma$ (these are the usual $\ell=1$ spherical harmonics) and $dv_\sigma$ is the area form of the metric $\sigma$. The positivity of the Bondi mass

 \begin{equation}\label{pmt} \sqrt{e^2-\sum_i p_i^2}\end{equation} was proved by Schoen-Yau \cite{SY} and Horowitz-Perry \cite{HP} under the dominant energy condition and a global assumption on horizon, see also \cite{CJS} and \cite{HYZ}. The supplementary equations imply the following equation satisfied by the mass aspect along $\mathscr{I}^+$:
\begin{equation}\label{du_mass_aspect}\partial_u m=\frac{1}{4}\nabla^a\nabla^b (\partial_u C_{ab})-\frac{1}{8}|\partial_u C|_\sigma^2,\end{equation} where $|\partial_u C|_\sigma^2=\sigma^{ac}\sigma^{bd} \partial_u C_{ab}\partial_u C_{cd}$. 
Integrating over $S^2_\infty$ with the metric $\sigma$ yields the well-known Bondi mass loss formula:
\begin{equation}\label{energy_loss1}\frac{d}{du} e(u)=-\frac{1}{32\pi}\int_{S^2_\infty} |\partial_u C|_\sigma^2 dv_\sigma \leq 0.\end{equation} In particular,
\begin{equation} \label{energy_loss2} e(u_1)\leq e(u_0) \text{ if } u_1\geq u_0.\end{equation}

%\begin{center}
%{\includegraphics[height=5cm]{P3.JPG}}
%\end{center}

This formula indeed corresponds to energy loss, see \cite{HYZ} for a monotonicity formula for the quantity $e-\sqrt{\sum_i p_i^2}$.

\subsection{The equivariance of Bondi mass under the BMS group}
Rescaling the spacetime metric \eqref{spacetime_metric} by $r^{-2}$ as $r\rightarrow \infty$, the limit of $r^{-2}g_{\alpha\beta} dx^\alpha dx^\beta$ approaches  $ \sigma_{ab} dx^a dx^b$, or the null metric on $\mathscr{I}^+$.\footnote{This is a special case of conformal compactification. In general, the metric on the unphysical spacetime is of the form $\Omega^2 g_{\alpha\beta} dx^\alpha dx^\beta$ and $\Omega=0$ corresponds to $\mathscr{I}^+$, see \cite{Penrose3, Penrose4, Geroch}.}  Therefore, $\mathscr{I}^+$ can be view as a null three-manifold:  
\[ \mathscr{I}^+= I\times (S^2, \sigma_{ab})\] with $u\in I$, $x^a\in S^2$. 

Each spacetime Bondi-Sachs coordinate system $(u, r, x^a)$ induces such a limiting coordinate system $(u, x^a)$ on $\mathscr{I}^+$, together with the mass aspect $m(u, x^a)$ and the shear $C_{ab}(u, x^a)$. Such a Bondi-Sachs coordinate system is by no means unique and the BMS group, which corresponds to the diffeomorphism group that preserves the gauge and boundary conditions, acts on the set of Bondi-Sachs coordinate systems. 

 A BMS group element induces a diffeomorphism $\mathfrak{g}$ on $\mathscr{I}^+$ that is of the following form:
\begin{equation}\label{bms1} \mathfrak{g}: (u, x^a)\mapsto (\bar{u}, \bar{x}^A), a=2, 3, A=2, 3\end{equation}
such that \begin{equation}\label{bms2} \begin{cases} \bar{x}^A&=g^A(x^a)\\ \bar{u}&=K(x^a)(u+f(x^a))\end{cases}\end{equation}
where $g: (S^2, \sigma)\rightarrow (S^2, \bar{\sigma})$ is a conformal isometry, i.e. $g^*\bar{\sigma}=K^2 \sigma$ where $K=(\alpha_0+\alpha_i Y_i)^{-1}$ and $(\alpha_0, \alpha_i)$ is a future timelike unit vector. 

Here is how the Poincar\`e group sits in the BMS group:

(1)  $f(x^a)$ is any smooth function on $S^2$ that is called a ``supertranslation".  $f(x^a)=\sum a_i Y_i$ corresponds to an actual translation in the Poincar\`e group.

(2)  $K=(\alpha_0+\sum \alpha_i Y_i)^{-1}$ corresponds to boosts in $O(3, 1)$. 

(3) Choices of $Y_i, i=1, 2, 3$ correspond to $O(3)\subset O(3,1)$.

The invariance/equivariance of the Bondi-Sachs energy-momentum is best described in terms of the modified mass aspect 2-form \cite{MTW}:

 \begin{definition}\label{modified_mass}
The modified mass aspect 2-form $\mathfrak{m}$ of a limiting Bondi-Sachs coordinate system $(u, x^a, \sigma)$ of $\mathscr{I}^+$ is defined to be
\[\mathfrak{m}=\widehat{m} dv_\sigma,\] where $\widehat{m}$ is 
\begin{equation}\label{mma}\widehat{m}=m-\frac{1}{4}\nabla^a\nabla^b C_{ab}, \end{equation} $\nabla$ is the covariant derivative with respect to the metric $\sigma$,
and $dv_\sigma=\sqrt{\det \sigma} dx^2\wedge dx^3$ is the volume form of the Riemannian metric $\sigma$. 
\end{definition}

Let $\mathfrak{m}$ and $\bar{\mathfrak{m}}$ be the modified mass aspects of the limiting Bond-Sachs coordinate systems $(u, x^a)$ and $(\bar{u}, \bar{x}^A)$ on $\mathscr{I}^+$, respectively. Suppose $(u, x^a)$ and $(\bar{u}, \bar{x}^A)$ are related by a BMS element $(K, f)$ as in \eqref{bms2}. The two modified mass aspect 2-form are related by
\begin{equation}\label{transformation_form}K^{-1}(\mathfrak{m}-\frac{1}{4} \Delta(\Delta+2) f dv_\sigma)=\mathfrak{g}^* \bar{\mathfrak{m}},\end{equation} where $\bar{\mathfrak{m}}=\widehat{\bar{m}} dv_{\bar{\sigma}}$ is the modified mass aspect 2-form of the limiting Bondi-Sachs coordinate system $(\bar{u}, \bar{x}^A, \bar{\sigma})$. 
This formula shows the equivariance of the Bondi-Sachs energy-momentum.

\begin{proposition} \cite{MTW} For any section $\Sigma$ of $\mathscr{I}^+$, suppose $\mathfrak{m}$ and $\bar{\mathfrak{m}}$ are the mass aspect 2-forms of two Bondi-Sachs coordinate systems which are related by a BMS group element that is a pure supertranslation, then the energy integrals are the same \[\int_\Sigma \mathfrak{m}=\int_\Sigma \bar{\mathfrak{m}}.\]
In presence of a nontrivial $K$, the energy-momentum transforms in the following way: 

Suppose  $\{Y_i\}_{i=1, 2, 3}$ be an orthonormal basis of the $(-2)$ eigenspace of $\Delta_\sigma$ and $K=(\alpha_0+\sum_j \alpha_j Y_j)^{-1}$ for a future timelike unit vector $(\alpha_0, \alpha_j)$. 
Suppose $A^\alpha_\beta\in O(3,1)$ satisfies $A_0^0=\alpha_0, A_0^k=\alpha_k$ and let $\bar{Y}_i= (A_i^0+A_i^k Y_k)K, i=1, 2, 3$. 
Denote \[ {e}=\int_\Sigma \mathfrak{m} , \,\, {p}_i=\int_\Sigma Y_i \mathfrak{m},\text{ and } \bar{e}=\int_\Sigma \bar{\mathfrak{m}},\,\, \bar{p}_i=\int_\Sigma \bar{Y}_i  
\bar{\mathfrak{m}}.\]   Then
\[\bar{e}=A_0^0e+A_0^k p_k,    \text{ and }    \bar{p}_i= A^0_i e+A_i^k p_k.\] In particular, $\bar{e}^2-\sum \bar{p}_i^2=e^2-\sum p_i^2$. 

\end{proposition}

We note that in this formulation, $\Sigma$ does not need to be the level set of any Bondi-Sachs coordinate $u$ on $\mathscr{I}^+$.

The energy loss formula \eqref{energy_loss2} can also be extended by means of the modified mass aspect two-form. 

\begin{definition}For any two sections $\Sigma_1$ and $\Sigma_2$ on $\mathscr{I}^+$, $\Sigma_1$ is said to be in the retarded future of $\Sigma_2$ if there exists a limiting Bondi-Sachs coordinate system $(u, x^a)$ such that $\Sigma_1$ and $\Sigma_2$ are given by $u=h_1 (x^a)$ and $u=h_2 (x^a)$ respectively, and that $h_1 (x^a)\geq h_2 (x^a)$ for each $x^a\in S^2$. 
\end{definition}
One easily check that this notion is independent of the choice of the limiting Bondi-Sachs coordinate system because $K>0$ and \eqref{bms2}. 

\begin{theorem} For any two sections $\Sigma_1$ and $\Sigma_2$ on $\mathscr{I}^+$ such that $\Sigma_1$ is in the retarded future of $\Sigma_2$, we have 
\[\int_{\Sigma_1} \mathfrak{m}\leq \int_{\Sigma_2} \mathfrak{m}.\]
 
\end{theorem}

Equation \eqref{du_mass_aspect} implies that the mass aspect 2-form $\mathfrak{m}$, as a 2-form on the three-manifold $\mathscr{I}^+$, verifies \begin{equation}\label{exterior_d} d\mathfrak{m}=-\frac{1}{8} |\partial_u C |_\sigma^2 du\wedge dv_\sigma , \end{equation} where $d$ is the exterior derivative operator on $\mathscr{I}^+$ as a differentiable manifold. 

This theorem should be considered as an extension of the classical Bondi mass loss formula \eqref{energy_loss2} which only applies to the case when $\Sigma_1$ and $\Sigma_2$  are both smooth and $u$ level sets of a fixed Bondi-Sachs coordinate system.

%%%%%%%%%%%%%%%%%%%%%%%%%%%%%%%%%%%%%%%%%%%%%%%%%%%%%%%%%%%

\subsection{Definitions of angular momentum at null infinity}

 There is also an angular momentum aspect $N^a$ which appears in further expansions of $W^a$ in \eqref{spacetime_metric} (we follow the convention in \cite{KWY}):
\[ W^a=\frac{1}{2r^2} \nabla^b C_{ab}+r^{-3} \left(\frac{2}{3}N^a-\frac{1}{16}\nabla^a(C_{de} C^{de})-\frac{1}{2} C_b^{\,\,a} \nabla^d C_d^{\,\, b}\right)+O(r^{-4}).\]

%Similar to \eqref{du_mass_aspect}, the vacuum Einstein equation implies 
%\[ 3\frac{\partial N_a}{\partial u}=-\nabla_a m+\frac{1}{4} \epsilon_a^b\nabla_b(\epsilon^{ec}\nabla_c \nabla^d C_{de})-\frac{3}{4} C_{ab} \nabla_c N^{bc}+\frac{1}{4} N^{cd} \nabla_d C_{ac}.\]

 All existing definitions of angular momentum at $\mathscr{I}^+$ includes the term $\int_{S^2_\infty} Y^a N_a$ where $Y^a$ is a rotation Killing field.  Different approaches based on Hamiltonian/spinor-twistor with respect to the BMS algebra lead to different definitions. For example, the expression in \cite{HPS} is 
 $N^a-u \mathring{\nabla}^a m$,  while the experssion in \cite{BT} is $N^a-\frac{3}{16} C_{bd} \mathring{\nabla}^a C^{bd}- \frac{1}{4} C^{ab}\mathring{\nabla}^d C_{db}$.

Some key issues that need to deal with for a definition of angular momentum at null infinity are: 1. Referencing, For example, the definition should be zero for the Minkowski time. 
2. Supertranslation ambiguity and Lorentzian ambiguity in the BMS group.

The calculation of the limit of the Chen-Wang-Yau quasilocal conserved quantities in Bondi-Sachs coordinates was taken up  by Keller-Wang-Yau in  \cite{KWY}. The expression depends on the Hodge decomposition of $C_{ab}$. Write
\[ C_{ab}=\nabla_a\nabla_b {\bf c}-\frac{1}{2}\sigma_{ab} \Delta {\bf c}+\frac{1}{2}(\epsilon_{ad} \nabla^d \nabla_b \underline{\bf c}+\epsilon_{bd} \nabla^d \nabla_a \underline{\bf c})\] and the limit of the CWY angular momentum (assuming the linear momentum vanishes) is 
\begin{equation}\label{KWY}\frac{1}{8\pi} \int_{S^2} Y_a\left(N^a-{\bf c}\mathring{\nabla}^a m  -\frac{1}{4} C^{ab}\mathring{\nabla}^d C_{db}\right).\end{equation}

A natural question is whether there exist a modified angular momentum aspect that satisfies similar properties of the modified mass aspect Definition \ref{modified_mass}, especially the equivariance property under the BMS group. Unfortunately, the transformation of the angular momentum aspect is extremely complicated. Chru\'sciel-Jezierski-Kijowski \cite{CJK}, through the Hamiltonian theory associated with Bondi-Sachs coordinates, defined the total Lorentz charge and showed that the Lorentz charge is equivariant under the BMS group if there exists a Bondi-Sachs coordinate system such that the mass aspect $m$ is a constant, $N_a$ is a parallel 1-form, and $C_{ab}=0$ (this corresponds to a stationary spacetime assumption).
The expression for angular momentum in \cite{CJK} is 
\begin{equation}\label{CJK}\frac{1}{8\pi} \int_{S^2} Y_a\left(N^a -\frac{1}{4} C^{ab}\mathring{\nabla}^d C_{db}\right).\end{equation}

Comparison of \eqref{KWY} and \eqref{CJK} shows that the expression from the limit of Chen-Wang-Yau quasilocal angular momentum satisfies the same equivariance property under the BMS group for the type of stationary spacetime considered in \eqref{CJK}.

 All previous definitions of angular momentum on $\mathfrak{I}^+$ depend on a specific gauge (a null frame or a spacetime coordinate system). In contrast, the Chen-Wang-Yau definition is geometric and coordinate independent (it depends only on $\sigma, {\bf H}$). In addition, solving the optimal isometric equation is a canonical procedure that is free from any ad hoc referencing or normalization. We thus expect more invariance/equivariance to hold true for this definition.

 In an upcoming paper, we compute the limit of the Chen-Wang-Yau quasilocal angular momentum on a general null hypersurfaces with the following goals:
 
 (1) Remove the determinant condition or the outgoing radiation condition in Bondi-Sachs coordinate system 
 
 (2) Extend the definition of angular momentum to a general section of $\mathscr{I}^+$ of the form $u=h(x)$. All calculations done in the Bondi-Sachs coordinate system so far work only for section of the form $u= constant$. 
 
 (3) Remove the condition of vanishing linear momentum in \cite{KWY}.
 
 We believe the result will get us closer to an angular momentum definition that satisfies stronger BMS equivarance property and to a better understanding of angular momentum after gravitational radiation.

\end{document}